# THE PHYSICAL NATURE OF FIVE-DIMENSIONAL SOLITONS: A SURVEY

Paul S. Wesson

1. Department of Physics and Astronomy, University of Waterloo, Waterloo, Ontario N2L 3G1, Canada.

2. Herzberg Institute of Astrophysics, National Research Council, Victoria, B.C. V9E 2E7, Canada.

Abstract: The basic quasi-Schwarzschild 5D objects known as solitons have a long history, which is reviewed. Then some material is added, leading to the inference that a soliton is a singularity in the geometry which represents a bivalent source of gravitational and scalar mass.

Correspondence: mail to (1) above; email = psw.papers@yahoo.ca

THE PHYSICAL NATURE OF FIVE-DIMENSIONAL SOLITONS: A SURVEY

1. Introduction

Solitons are five-dimensional objects whose metric is static, spherically symmetric in ordinary space and asymptotically flat. These properties are shared by the unique Schwarzschild solution of four-dimensional general relativity. However, it is a mistake to confuse the two. We wish to clear up this and other misunderstandings about the physical nature of solitons, and will conclude that in general they possess gravitational and scalar 'charges' which can be regarded as bivalent aspects of mass.

Section 1 gives a short summary of the long and convoluted history of the solitons, to establish what is already known about them. Section 2 focuses on the physical properties of solitons, and offers some insights about their real nature. Section 3 is a conclusion. The notation is standard: indices $A$, $B$ run over 5D while $\alpha, \beta$ run over 4D, and the physical constants are absorbed unless they are made explicit to assist interpretation.

2. The Saga of the Solitons

5D relativity is a unified theory of tensor, vector and scalar interactions [1]. These are normally identified with gravity, electromagnetism and a field whose nature is obscure but may be connected with inertial mass. Metrics of the theory are solutions of field equations which are commonly specified by setting the 5D Ricci tensor to zero: $R_{AB} = 0$ where $A, B = 0, 123, 4$ for time, space and the extra dimension. The soliton solutions may resemble higher-dimensional one-body fields, but they have a long and contentious history.

What is now known as the 5D soliton metric was first obtained in 1970 as an extension of the 4D Schwarzschild one by Kramer [2]. It was later derived independently by Sorkin [3], who was convinced it described a magnetic monopole, an interpretation adopted by Gross and Perry [4], who however also realized that it involved two kinds of mass. Then Davidson and Owen derived it under the belief it described a kind of black hole [5], a view criticized by Dereli [6]. Related metrics were studied by Sokolowski and Carr [7] and Liu [8], and a version of the soliton was used to model a wormhole by Agnese *et al*. [9]. The soliton metric or special cases of it were 'discovered' again by other workers [10, 11]; but there was no consensus about the physical meaning of the metric.

This situation changed significantly in 1992, when Wesson calculated an effective source for the soliton metric [12]. His method was also applied to cosmology [13], and developed with Ponce de Leon into a general scheme to derive the 4D physical source for any 5D metric [14, 15]. Together, these workers applied the Tolman-Whittaker integral to calculate the effective 4D gravitational mass of a soliton [16]. This is the source which determines the orbit of a test particle in the field of the soliton.

By contrast, what is commonly termed the 'energy' of the soliton may be calculated using only the metric, without regard to any matter that may be associated with it. This metric-based energy was considered by Gross and Perry [4] and discussed at length by Bombelli *et al*. [17]; but a later calculation by Deser and Soldate [18] gave a slightly discrepant answer. The subject was taken up again by Billyard *et al*. [19], who confirmed the earlier result. It was further confirmed by Sajko and Wesson [20, 21], who used a Hamiltonian method to calculate the metric-based energy of electrically-charged solitons.

The extension of the soliton class of metrics to include electric charge had been made slightly earlier [22, 23]. However, it should be mentioned that a different approach to charged 5D objects had been made in 1985 by Nodvik [24], who in a little-noticed paper argued that such objects could cause 4D vacuum polarization. This subject remains controversial, but should be examined in detail because Liko [25] has shown that 5D solutions can have measurable electromagnetic effects, and it is currently a question as to whether the fine-structure constant varies with cosmological time.

There are other controversial aspects of 5D relativity, especially in regard to the basic (uncharged) solitons. Ponce de Leon has made intensive efforts to resolve some of these. He has related the solitons to an extra symmetry of the (3D) spherically-symmetric field equations [26], looked at the status of the Weak Equivalence Principle in 5D [27], and made a detailed investigation of the two parameters which define the soliton class of solutions [28]. The last study agrees for the most part with an investigation of the geometrical properties of solitons by Lake [29]. These recent investigations are mainly concerned with theoretical properties of the 5D soliton class of solutions, and in differentiating it from the 4D Schwarzschild solution.

Experimental information about the solitons should also be mentioned, even though it dates from the time when they were thought to be matter-free extensions of the unique 4D solution. A comprehensive comparison with the classical solar-system tests was made by Kalligas *et al.* [30]. This showed that the two soliton parameters must be very close to the values which give back the Schwarzschild solution. The two parameters concerned obey a quadratic consistency relation which will be examined in detail in Section 3. All researchers agree that this consistency relation is crucial to an understanding of the solitons, but opinions differ as to how it should be most usefully presented. In an alternative investigation of the solar-system tests, Lim *et al.* [31] suggested that the relation could be cast in the form of the equation for an ellipse. This possibility was made explicit by Ponce de Leon [28], and a different version of this will be presented below. In regard to experimental work on the solitons, it was suggested that the classical tests could be augmented by investigating the precession of a spinning object in Earth orbit, as in the Gravity-Probe B or Stanford gyroscope experiment. The basic relations for this were worked out by Liu and Wesson [32], but the magnitude of the difference from general relativity was too small to be detected by that experiment. More general studies of this effect and the other solar-system tests were made by Liu and Overduin [33, 34]. They showed that in a comparison of the 5D soliton metric with the 4D Schwarzschild metric, the bodies of the solar system agree with the latter to high accuracy, in some cases better than 1 part in $10^6$.

This level of support for the 4D theory over the 5D one is exceptional, and made some workers suspicious that there must be some special reason for it. This was provided in 1994 by Mashhoon *et al.* [35], who found a new 5D embedding for the 4D Schwarzschild solution. The so-called canonical embedding actually leads to dynamical results in 5D and 4D which are *identical*. This can be traced to the fact that the canonical metric, unlike the soliton one, has a flat extra dimension and a quadratic prefactor in the extra coordinate which multiplies onto 4D spacetime. Indeed, it is a theorem that the 5D canonical metric smoothly embeds *any* 4D vacuum metric of general relativity. This strong result follows from the broader embedding theorem of Campbell, which was formulated in the 1920s when Kaluza and Klein were building the foundations of 5D relativity. However, Campbell's theorem lay forgotten in the literature, until Tavakol and coworkers drew attention to it in the 1990s [36-38]. The theorem also implies that the 5D field equations in apparent vacuum $R_{AB} = 0 \,(A, B = 0-5)$ *contain* the 4D Einstein equations with effective matter $G_{\alpha\beta} = 8\pi T_{\alpha\beta} \,(\alpha, \beta = 0-3)$. Campbell's theorem put a solid geometrical planck under the previously physical theory, which came to be known as induced-matter or space-time-matter theory (see ref. 39 for a recent review). The effective or induced matter for the solitons will be specified below. Here, it can be noted that the alternative current version of noncompactified Kaluza-Klein relativity, namely membrane theory, employs a singular hypersurface in the 5D manifold to identify spacetime, but is mathematically similar to space-time-matter theory. Campbell's theorem can be applied to many problems in 5D relativity [40], and considerably simplifies the interpretation of solutions of the theory.

Another simplification, which affects the dynamics of the theory, is that the 4D paths of all particles (including massive ones) are equivalent to 5D *null* paths. As a working hypothesis, this was suggested in connection with space-time-matter theory [41, 42], but it also applies to membrane theory [43]. The null-path hypothesis is especially useful when applied to metrics with a high degree of symmetry, such as those in cosmology. In this context, it should be mentioned that there is a time-dependent version of the soliton metric [44], in which the static soliton evolves into one with a Hubble-like expansion, suggesting that solitons may be viable models for perturbations in the early universe. In general, the 5D null-path hypothesis (which preserves 4D causality) provides an embedding for the dynamics, while Campbell's theorem provides an embedding for the solutions, so there is a smooth transition from Kaluza-Klein theory to Einstein theory.

3.  <u>The Physical Nature of the Solitons</u>

Despite the extensive work on solitons as reviewed in the preceding section, their physical nature remains controversial. Why is this?

A major reason concerns the relative standing of the 4D theory and the 5D theory. Are they separate, or is one contained in the other? Recent work strongly favours the latter belief. It then follows by Campbell's theorem that the solitons are not vacuum solutions (except in one very special case of the two defining parameters). Rather, while the 5D field equations appear to be empty, they actually involve induced or effective matter which is necessary to balance the 4D Einstein equations. That is, while the solutions are found from $R_{AB} = 0$, they also satisfy $G_{\alpha\beta} = 8\pi T_{\alpha\beta}$ with a finite energy-momentum tensor. This is indisputably true as a statement about algebra, but it implies a shift in view about physics.

A further source of confusion about the relative standing of the 4D and 5D theories concerns covariance. The 4D theory is covariant under the change $x^\alpha \to \bar{x}^\alpha (x^\beta)$, while the 5D theory is covariant under $x^A \to \bar{x}^A(x^B)$. The second group is broader than the first, so a gauge change which involves $x^4$ can lead to different metrics in $x^\alpha$, and potentially to different 4D physics. This has been demonstrated explicitly by the thorough work of Ponce de Leon [26-28] and others. While this flexibility does not directly affect the $x^4$-independent solitons, it does explain why there is another viable embedding for the Schwarzschild solution, namely the $x^4$-dependent canonical solution of Mashhoon *et al.* [35]. The implication is that the 'correct' 5D analog of the 4D Schwarzschild metric is not the soliton metric but the canonical metric. (Both will be given below, and a detailed discussion of the situation is given in ref. 1.) The identification of the canonical metric as the 1-body solution relevant to the field of the Sun also explains the high accuracy of the solar-system tests of 5D gravity.

A related subject concerns the choice of coordinates in 5D, and how this relates to distinguishing (or not) the gravitational field from the electromagnetic and scalar fields. In 5D there are obviously 5 coordinate degrees of freedom which can be used to simplify the geometry. For example, the canonical metric uses the 5 coordinate choices to set the components of the 5D metric tensor via $g_{4\alpha} = 0, |g_{44}| = 1$. This by the conventional interpretation removes the electromagnetic potentials and flattens the scalar potential, leaving the gravitational potentials $g_{\alpha\beta}(x^\gamma, x^4)$. However, this algebraic procedure in the general case can lead to problems of physical interpretation, because it is conceivable that effects of the other fields may be 'forced' into the gravitational sector. This situation does not arise in general relativity, since it deals solely with gravity. (In the Reissner-Nordstrom solution, the electrical charge on the central mass only enters via its gravitational energy.) But it is problematic in any *classical* theory of the Kaluza-Klein type, since the *quantum* information necessary to separate the fields is lacking. (The graviton has spin 2, the photon has spin 1 and the scaleron has spin 0, so the putative particles which make up the fields can be distinguished, for example by their polarization states.) This problem is exacerbated in classical 5D theories by the fact that the potentials in the appropriate limits all vary with distance as $1/r$, reflecting the massless nature of the corresponding quanta. A quantum Kaluza-Klein theory would not be subject to this problem; but in the absence of a quantum version of Einstein's theory it is not known what form a quantum unified theory would take. For the present, there appears to be no option but to assign the physical nature of the fields according to where their sources appear in the metric.

By this rule, it is possible to immediately dismiss the idea that the source for the soliton metric is electromagnetic in nature. This because the soliton metric is diagonal (see below), while the electromagnetic potentials $A_\mu$ correspond to off-diagonal components of the metric tensor. The most commonly used metric for this type of problem has a 5D line element which contains the 4D one thus:

$$dS^2 = ds^2 + \varepsilon \Phi^2 \left( dx^4 + A_\mu dx^\mu \right)^2 \quad . \tag{1}$$

Here $\varepsilon = \pm 1$ determines whether the extra dimension is spacelike or timelike, the latter being allowed since the extra coordinate in modern Kaluza-Klein theory does not have the physical nature of a time, so there are no contradictions of conventional causality $\left(ds^2 \geq 0\right)$. The scalar potential $\Phi$ can in principle depend on all 5 coordinates $\left(x^\gamma, x^4\right)$; and because of the way in which the potentials are defined, the electromagnetic and scalar fields may be mixed in modern 5D relativity [15, 24]. Also, if the electric charge and the mass of the source are $Q$ and $M$, in the 5D charged-soliton solutions [22, 23] the combination $Q^2/Mr$ occurs, whereas $Q^2/r^2$ appears in the 4D Reissner-Nordstrom solution. The existence of the charged solitons, which in the neutral limit give back the standard solitons, is itself an argument against the source for the latter being a magnetic monopole or an isolated electric charge. (As noted by Sorkin in refs. 3 and 17, it is really arbitrary which of these is chosen.) Also, it should be recalled that while they were predicted by Dirac from his theory of the electron, magnetic monopoles have been searched for assiduously, with zero success.

Given that the standard solitons do not involve electromagnetism, there remains gravitational mass and scalar charge as possible sources. But as mentioned above, the electromagnetic $E$, gravitational $G$ and scalar $S$ fields all behave in the same way, at least for large distances $r$ from the centre of the geometry. Some ambiguity is therefore to be expected in separating the $G$ and $S$ fields. This difficulty is compounded by the fact that two different coordinate systems have been used in studying solitons: spatially isotropic coordinates are best suited to calculating their effective matter; while quasi-Schwarzschild coordinates are best suited to separating the $G$ and $S$ fields and comparing these with experimental data. A further, though less significant, problem is that different workers have used different terminologies for the two dimensionless parameters which define the soliton class of metrics. These will be denoted $a$ and $b$ in what follows. Those readers who wish to translate the three older terminologies into the present one may refer to [45]. An advantage of the newer terminology is that it allows the *G* and *S* states to be fixed by finite values of the parameters, rather than by (in some cases) limits. The two parameters concerned satisfy a consistency relation set by the spatial part of the 5D field equations $(R_{ij} = 0, i \neq j)$. It reads

$$a^2 + ab + b^2 = 1 \quad . \tag{2}$$

This relation is invariant under a swap of *a* and *b*, and a change in the sign of both. These properties also follow from the field equations, and notably the extra symmetry in them discussed by Ponce de Leon [26, 46]. The consistency relation (2) is central to a proper understanding of the solitons.

The metric for the solitons in isotropic coordinates is given by

$$dS^2 = \left(\frac{1-M/2r}{1+M/2r}\right)^{2a} dt^2 - \left(\frac{1-M/2r}{1+M/2r}\right)^{2(1-a-b)} \left(1+\frac{M}{2r}\right)^4 \left[dr^2 + r^2 d\Omega^2\right]$$

$$- \left(\frac{1-M/2r}{1+M/2r}\right)^{2b} dl^2 \quad . \tag{3}$$

Here $d\Omega^2 \equiv \left(d\theta^2 + \sin^2\theta d\phi^2\right)$, and the extra dimension is taken to be spacelike. The constant *M* is used for reasons of convention, but as will become clear below it should *not* be identified with a simple mass at the centre of the geometry, as in the 4D Schwarzschild solution. The latter, plus an extra flat piece, is recovered from (3) when $a = 1$, $b = 0$. Also, the centre of the spatial geometry is now at $r = M/2$, and the temporal part of the geometry goes to zero at the same place. This means that the solitons in general do not have horizons of the conventional sort, as remarked by several workers. That is, the solitons should not be called black holes.

Furthermore, the solitons have matter associated with them. To see this, recall that they are solutions of the 5D field equations $R_{AB} = 0$, and if they are also to be solutions of the 4D Einstein equations $G_{\alpha\beta} = 8\pi T_{\alpha\beta}$ then the former set of equations has to be systematically reduced to the latter set. The general method for doing this has by now become standard since its introduction in 1992 [14; 13, 12]. Basically, the 5D $R_{AB}$ is decomposed into the 4D $R_{\alpha\beta}$ plus another part which depends on the extra dimension, in the form of terms which involve derivatives with respect to the extra coordinate and the extra metric coefficient. The 4D Ricci scalar $R$ is obtained in similar form by contraction. The 4D Einstein tensor $G_{\alpha\beta} \equiv R_{\alpha\beta} - (R/2) g_{\alpha\beta}$ is then constructed, and matched to an effective $T_{\alpha\beta}$ which is 'induced' by the extra dimension. The result is that $R_{AB} = 0$ also imply $G_{\alpha\beta} = 8\pi T_{\alpha\beta}$ where $T_{\alpha\beta} = T_{\alpha\beta}(\partial g_{\mu\nu} / \partial x^4, g_{44})$. The $T_{\alpha\beta}$ calculated in this manner from the metric (3) only depends on the $r$ coordinate and the constants $M$, $a$ and $b$, where the latter are constrained by the consistency relation (2). Specifically, the density and the components of the (anisotropic) pressure are given by

$$8\pi\rho = \frac{-abM^2 r^4}{(r-M/2)^4 (r+M/2)^4} \left(\frac{r-M/2}{r+M/2}\right)^{2(a+b)} \tag{4.1}$$

$$8\pi p_r = \frac{2bMr^3}{(r-M/2)^3(r+M/2)^3}\left(\frac{r-M/2}{r+M/2}\right)^{2(a+b)}$$

$$+ \frac{bM^2r^4(a+2b-4r/M)}{(r-M/2)^4(r+M/2)^4}\left(\frac{r-M/2}{r+M/2}\right)^{2(a+b)} \quad (4.2)$$

$$8\pi p_{\theta,\phi} = \frac{-bMr^3}{(r-M/2)^3(r+M/2)^3}\left(\frac{r-M/2}{r+M/2}\right)^{2(a+b)}$$

$$\frac{-bM^2r^4(a+b-2r/M)}{(1-M/r)^4(1+M/r)^4}\left(\frac{r-M/2}{r+M/2}\right)^{2(a+b)} \quad . \quad (4.3)$$

These components satisfy the equation of state for radiation or ultrarelativistic particles. Defining $\bar{p} \equiv (p_r + p_\theta + p_\phi)/3$, this is $\bar{p} = \rho/3$. It is in fact a theorem that for 5D metrics independent of $x^4$ the components of $T_\beta^\alpha$ satisfy

$$T \equiv T_\alpha^\alpha = T_0^0 + T_1^1 + T_2^2 + T_3^3 = \rho - p_r - p_\theta - p_\phi = 0 \quad . \quad (5)$$

For $T_0^0 = \rho > 0$, (4.1) above shows that the dimensionless parameters $a$ and $b$ must have opposite signs.

The gravitational mass of the fluid defined by (4) is given by the Tolman-Whittaker formula mentioned before, which is an appropriately-defined integral over $(T_0^0 - T_1^1 - T_2^2 - T_3^3) = (\rho + p_r + p_\theta + p_\phi)$. It is

$$M_g(r) = aM\left(\frac{r-M/2}{r+M/2}\right)^{-b} \quad . \quad (6)$$

For this to be positive requires $a > 0$, so by the preceding condition for positive density it is necessary that $b < 0$. Then the exponent in (6) is positive, and $M_g(r) \to 0$ for $r \to M/2$. This is the centre of the geometry as defined above, so the gravitational mass is zero at the physical centre of ordinary space. Also $M_g(r) \to aM$ for $r \to \infty$, so the asymptotic 5D mass is in general not equal to the standard 4D mass. The Schwarzschild case (a=1, b=0) is exceptional, and from the viewpoint of functional analysis not really a member of the soliton class. In general, solitons are spherically-symmetric clouds of radiation-like matter whose properties fall off rapidly at distance $(\sim 1/r^4)$ but extend indefinitely far into space.

The gravitational mass $M_g(r)$ of (6) is obtained from an integral over what in cosmology is called the gravitational density $(\rho + 3p)$. This is the appropriate density for calculating the attraction of a portion of a continuous fluid in accordance with the radial equation of motion (also known in this application as the Raychaudhuri equation). There is also the so-called inertial density $(\rho + p)$, which governs the mechanical properties of matter in accordance with the equation of continuity. For radiation-like matter, the equation of state is given by $T = 0$ or $\bar{p} = \rho/3$. Then the gravitational and inertial densities are $2\rho$ and $4\rho/3$ respectively. This means that for the solitons, their effective inertial mass is $M_i(r) = (2/3)M_g(r)$. Insofar as the Weak Equivalence Principle is sometimes loosely stated as the proportionality of gravitational and inertial mass, the solitons obey it. Of course, most ordinary matter in the universe (as opposed to radiation or vacuum) has $p \ll \rho$, so the gravitational and inertial masses are equal to a good approximation. A more sophisticated discussion of 5D objects and the Weak Equivalence

Principle has been given by Ponce de Leon [27]; and the astrophysical implications of solitons have been discussed by Wesson [47]. They could be relevant to dark matter and pre-galactic perturbations.

The nature of solitons becomes clearer under a change from isotropic to quasi-Schwarzschild coordinates via $r \to \left(r - M + \sqrt{r(r-2M)}\right)/2$. Then (3) becomes

$$dS^2 = \left(1-\frac{2M}{r}\right)^a dt^2 - \frac{dr^2}{(1-2M/r)^{a+b}} - \frac{r^2 d\Omega^2}{(1-2M/r)^{a+b-1}} - \left(1-\frac{2M}{r}\right)^b dl^2 \quad . \tag{7}$$

From this, it is obvious that for $(2M/r) \ll 1$ what would normally be called *the* mass is split into parts $aM$ and $bM$, which figure in the first and last terms and are frequently called the gravitational mass and the inertial mass. The former appellation is justified in that $M_g(r) \to aM$ for $r \gg M/2$ in (7). But the inertial mass of a source means different things to different people, so it is better to refer to $bM$ as the scalar mass.

The contending influences of the parameters $a$ and $b$ may be elucidated by considering the hypothetical situation where a test particle is momentarily as rest in 3D, has velocities normalized via $u^\alpha u_\alpha = 1$ in 4D and pursues a null-path in 5D (see Section 2). The relevant Christoffel symbols and the acceleration in the radial direction in the weak-field limit are given by

$$\Gamma^1_{00} = \frac{aM}{r^2}\left(1-\frac{2M}{r}\right)^{2a+b-1} \quad \Gamma^1_{44} = -\frac{bM}{r^2}\left(1-\frac{2M}{r}\right)^{a+2b-1}$$

$$\frac{d^2 r}{dt^2} = (-a+b)\frac{M}{r^2} \quad . \tag{8}$$

This shows that both the *G* and *S* fields contribute to the acceleration of a test particle (physicality requires $a > 0$, $b < 0$). The matter which causes this is specified by (4). That fluid has the equation of state $\bar{p} = \rho/3$; but it should not be automatically assumed that it consists of photons or an electromagnetic field, since the noted equation of state holds for any particles that are massless and move at the speed of light. Indeed, they could be gravitons and scalerons, since the metric (7) describes these kinds of fields.

The soliton metric (7) should not be confused with the Schwarzschild one for reasons already given. However, to make the situation clear it is desirable at this point to quote the 5D (pure) canonical metric, which provides an unambiguous embedding for the 4D Schwarzschild-deSitter solution:

$$dS^2 = \frac{\Lambda(x^4)^2}{3}\left[\left(1 - \frac{2M}{r} - \frac{\Lambda r^2}{3}\right)dt^2 - \frac{dr^2}{(1 - 2M/r - \Lambda r^2/3)} - r^2 d\Omega^2\right] - (dx^4)^2 \quad . \quad (9)$$

This metric, unlike that of the soliton, *does* describe a point mass at the centre of ordinary 3D space. The mass is embedded in a background with a cosmological constant $\Lambda$, which can be regarded as a vacuum fluid with the same equation of state as in Einstein's theory, namely $p_v = -\rho_v = -\Lambda/8\pi$. The quadratic dependence on the extra coordinate $x^4$ in (9) is characteristic of embeddings in 5D canonical space [35, 36], which should not be confused with embeddings in 5D Minkowski space [48]. 5D canonical spaces typically describe aspects of vacuum physics, and can be used as models for elementary particles [49]. By contrast, the soliton metrics (3) and (7) are asymptotically Minkowski and can be used as models for 'hot' astrophysical objects.

More information on the solitons can be obtained by plotting the consistency relation (2) for their defining parameters, as in Figure 1. Allowed values of $a$ and $b$ must lie on the tilted ellipse shown there. However, physical constraints $(a > 0, b < 0)$ mean that attention should be focused on the upper, left quadrant of that plot. The ellipse defined by (2) is tilted because of the cross-term in that relation.

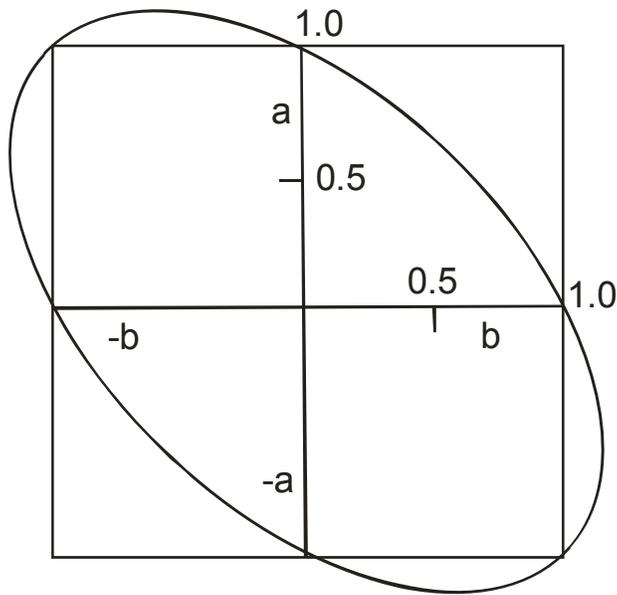

Figure 1.  The consistency relation $a^2 + ab + b^2 = 1$ for the 5D solitons, with a square of half-side unity for reference. The noted relation describes a tilted ellipse, with semi-major axis $\sqrt{2}$ and semi-minor axis $\sqrt{2/3}$.

The question arises of whether the ellipse can be rotated into its standard configuration. This can be accomplished by a coordinate transformation on the metric according to Dereli [6], or more directly by a redefinition of the parameters. It may be verified that the appropriate relations are:

$$\alpha = k_1 a + k_2 b \qquad \beta = k_2 a + k_1 b$$

$$a = \frac{\alpha k_1 - \beta k_2}{(k_1^2 - k_2^2)} \qquad b = \frac{\beta k_1 - \alpha k_2}{(k_1^2 - k_2^2)}$$

$$k_1 = \pm \left(\frac{1}{2} \pm \frac{\sqrt{3}}{4}\right)^{1/2} \qquad k_2 = \frac{1}{4}\left(\frac{1}{2} \pm \frac{\sqrt{3}}{4}\right)^{-1/2}. \qquad (10)$$

With the new parameters $\alpha$, $\beta$ in place of the old ones $a$, $b$ the consistency relation (2) reads

$$\alpha^2 + \beta^2 = 1 \quad . \qquad (11)$$

This is the equation of an ellipse in standard rectangular coordinates $x$, $y$ where $\alpha \equiv x/A$, $\beta \equiv y/B$ and $A$, $B$ are the semi-axes. Alternatively, the ellipse may be considered as the locus of the tip of a moving, right-angled triangle; and then (11) is just the statement that the sides (squared) sum to the hypotenuse (squared) with value unity. Physically, (11) means that the parameters defining the sources sum square-wise to unity. That is, in a sense the sources are parts of the same thing. Or, there is really only one source, which is bivalent.

This proves in a formal sense the view of several workers [6, 15, 26] that there is only one source $M$ which manifests itself as gravitational and scalar fields. The electromagnetic field is not manifest in the diagonal soliton metrics (3) and (7), but it can be included via the off-diagonal terms in (1). There, the electromagnetic potentials $A_\mu$ are essentially introduced by the coordinate shift $dx^4 \to dx^4 + A_\mu dx^\mu$, or equivalently by the velocity boost $u^4 \to u^4 + A_\mu u^\mu$. It is a reasonable conjecture that all three fields may be manifestations of a single source. In the conventional 4D treatment of electromagnetism, the source charge $Q$ produces an electric field $E = Q/4\pi\varepsilon_0 r^2$ (where $\varepsilon_0$ is the permittivity of free space) and this corresponds to an energy density $\rho \sim E^2 \sim Q^2/\varepsilon_0^2 r^4$ [1, 50]. It is interesting that the density of the soliton fluid (4.1) also varies asymptotically as $1/r^4$. If that fluid were electromagnetic in nature, the small-$r$ departures from Coulomb's law could be attributed to a change in the characteristics of the vacuum via $\varepsilon_0(r)$, which is basically the proposal of Nodvik [24]. As it is, the soliton fluid (4) is due to the gravitational and scalar fields, but the same principles can apply as follows: The large-$r$ behavior of the soliton density (4.1) is proportional to $(aGM)(bGM)/r^4$, where the geometry is approximately flat and $G$ is the conventional gravitational constant, which can be thought of as the analog of the electrical permittivity $(G \sim 1/\varepsilon_0)$; while the small-$r$ behaviour of the density is due to the non-flat geometry, which can be modeled in principle as a variation of the properties of the vacuum via $G(r)$. Irrespective of how the soliton fluid (4) is viewed, it is apparent that in general it is the result of a single, bivalent source.

This becomes even more obvious by writing $\alpha = \sin\theta$, $\beta = \cos\theta$ in the consistency relation (11), which then reads just $\sin^2\theta + \cos^2\theta = 1$. The use of an angle is the simplest way to parametize the solitons.

There remains the not-so-simple question of what, if anything, lies at the centre of the soliton geometry. Most workers have used the isotropic coordinates of (3), and have argued that nearly all solitons lack horizons because the physical centre lies at $r = M/2$. As this is approached, the density and pressure of (4) diverge, but the gravitational mass (6) goes to zero. Geometry is behind this strange behavior. The gravitational influence grows outwards, because more of the fluid is encompassed as the radius grows, and by Gauss' theorem this acts like a mass located at the centre. But there is no object of a gravitational nature at the actual 'centre' $r = M/2$. This point changes apparent location, however, when the coordinates are altered to the quasi-Schwarzschild ones of (7). Its location in the new coordinates is $r = 2M$, the horizon in the 4D 1-body solution. Mathematically, it may be tempting to explore the 'inner' part of the soliton metric; but physically it makes no sense, because the 5D Kretschmann scalar $K \equiv R_{ABCD}R^{ABCD}$ diverges at $r = M/2$ (isotropic coordinates) or $r = 2M$ (quasi-Schwarzschild coordinates). The 4D curvature scalar $C \equiv R_{\alpha\beta}R^{\alpha\beta}$ diverges at the same place [15, 16]. Since these are the geometric invariants widely acknowledged as indicating a singularity, it has to be concluded that the soliton geometry effectively ends at the centre as previously defined ($r = M/2$ or $r = 2M$ depending on the coordinates). Thus the centre of a soliton is a singularity, even though there is no gravitational mass residing there. This situation may be better appreciated by employing the old-fashioned notion of lines of force. These emerge from the singularity at the centre, where they are cramped, and diverge radially with ever widening spacing, and are responsible for the gravitational and scalar forces typical of the soliton. The matter specified by (4), which is necessary to balance Einstein's equations, is either a measure of the lines of force or of particles associated with them. Quantum information is needed to investigate this further, but the simplest interpretation is that the soliton matter consists of quanta of the fields (gravitons and scalerons). In loose language, the solitons resemble singular 'holes' in the geometry from which gravitational and scalar fields emerge.

4. Conclusion

The history of solitons is replete with discoveries and rediscoveries, mostly of a technical nature, and in Section 2 the main nuggets of knowledge were presented. The physical nature of solitons has a sparser inventory of reliable results, but the main features were discussed in Section 3 using the most logical terminology. The major properties of solitons can be summarized as follows:

(a) The soliton metric, in isotropic coordinates (3) or quasi-Schwarzschild coordinates (7), is not merely an extension to 5D of the 4D 1-body solution of general relativity. The true embedding of the 4D Schwarzschild metric is provided by the 5D canonical metric (9), which is especially relevant to a universe dominated by the cosmological constant.

(b) If the soliton solution of the 5D field equations $R_{AB} = 0$ is also to be a solution of the 4D Einstein equations $G_{\alpha\beta} = 8\pi T_{\alpha\beta}$, then matter must be present with density and pressure given by (4). The equation of state (5) of this matter is that of ultrarelativistic particles or massless quanta. Since the soliton lacks the off-diagonal terms typical of the 5D metric (1) with electromagnetism and photons (spin 1), it is reasonable to assume that the soliton field consists of gravitons (spin 2) and scalerons (spin 0). However, the gravitational mass (6) of the soliton, as calculated from the usual definition, goes to zero at the physical centre of the geometry.

(c) The gravitational and scalar fields of the soliton, typically related to the time and extra parts of the metric, can be shown by a new choice of parameters (10), (11) to be aspects of one source. In the simplest formulation, both are specified by an angle in a hypothetical 2D space. Physically, the single source shows gravitational and scalar characteristics, and is bivalent.

(d) Geometrically, the physical centre of the soliton metric is typified by unbounded values of the 5D and 4D invariants. While there is no gravitational mass located there, the centre is a singularity from which gravitational and scalar fields diverge.

The preceding comments show that solitons are indeed strange objects. They will probably continue to be rediscovered by eager if ill-read workers; but what is really needed is a better understanding of their physics, and perhaps an observational proof of their existence.

Acknowledgements

I am grateful for comments over the years by various members of the Space-Time-Matter consortium (http://astro.uwaterloo.ca/~wesson). This work was partially supported by N.S.E.R.C.